\documentclass[preprintnumbers,amsmath,amssymb,twocolumn]{revtex4}

\usepackage{dcolumn}% Align table columns on decimal point
\usepackage{graphicx}
\usepackage{subfig}
\usepackage{verbatim}
\usepackage{color} 
\usepackage{amsmath}
\usepackage{array}
\usepackage{multirow}

\newcommand{\inbox}[1]{\multicolumn{1}{|>{\centering}m{0.14\textwidth}|}{#1}}
\newcommand{\edit}[1]{}

\begin{document}

\title{Two-Dimensional Packing of Soft Particles and the Soft Generalized Thomson Problem}
\author{William L. Miller}
\author{Angelo Cacciuto}
\email{ac2822@columbia.edu}
\affiliation{Department of Chemistry, Columbia University \\3000 Broadway, New York, New York 10027}
%\date{\today}
\begin{abstract}
We perform numerical simulations of \edit{a model of} purely repulsive  
soft colloidal particles interacting via a generalized elastic potential and   
constrained to a two-dimensional plane and to the surface of a spherical shell. 
For the planar case, we compute the phase diagram in terms of the system's rescaled density and temperature.
We find that a large number of ordered phases becomes accessible at low temperatures as the density of the system 
increases, and we study systematically how structural variety depends on the functional shape 
of the pair potential.
For the spherical case, we revisit the generalized Thomson problem for small numbers of particles $N\leq12$ 
and identify, enumerate and compare the minimal energy polyhedra established by the location of the 
particles to those of the corresponding electrostatic system.

\end{abstract}
 
\maketitle
\section*{Introduction}
Understanding how nanocomponents spontaneously organize into complex macroscopic structures
is one of the great challenges in the field of soft matter today. Indeed, the ability to predict and control 
the phase behavior of a solution, given a set of components, may open the way to the development
of materials with novel optical, mechanical, and electronic properties. 
Despite much effort in this direction, the question of how to link
the physical character of the components, i.e. their bare shape and their interaction potential, 
to their phase behavior still remains unanswered. 

Of particular interest for optical applications are the crystalline phases. Until recently, it was believed that 
shape anisotropy and/or directional interactions where key elements for the formation of 
crystals with complex symmetries commonly observed  for instance in atomic solids. The study of
the phase behavior of components interacting with exotic (yet isotropic) potentials has 
proven this belief to be incorrect. For instance, Torquato \textit{et al.} have shown, using inverse optimization techniques, that it is possible to achieve
non-closed-packed structures such as simple cubic, hexagonal,  wurtzite and even diamond phases
with isotropic pair potentials  (see ref~\cite{torquato} for a review on the subject). 
Furthermore, spherical particles interacting via a hard core and a repulsive shoulder potential are known to organize into complex mesophases not unlike those found with diblock copolymers~\cite{glaser2}. 

One class of pair potentials that has recently attracted much attention  
is that describing the interactions between soft/deformable mesoparticles. 
Unlike  typical colloidal particles for which excluded volume interactions are strictly enforced via a hard-core or a Lennard-Jones potential,  complex mesoparticles such as charged or neutral star polymers, dendrimers or microgels present a more peculiar pair potential describing their volume interactions. 
Surprisingly, the simple relaxation of the constraint of mutual impenetrability between isotropic 
components gives access to several non-close-packed crystalline structures at high densities.
Given the complexity of these mesoparticles, their interactions are usually 
extracted via an explicit coarse-grained procedure to obtain ad-hoc effective pair potentials. 
What emerges is a great variety of nontrivial interactions, some of which allow for even 
complete overlap among the components, resulting in a very rich phenomenological behavior.
Remarkably, it is feasible to engineer interactions between star polymers or dendrimers
by controlling their overall chemical/topological properties~\cite{mladek}. 
   
The phase behavior of several systems adopting these exotic, but physically inspired, interactions
has been the subject of several publications~\cite{louis,likos0,likosA,likosB,denton,gottawald,pierleoni, Capone, bozorgui,xu,prestipino}. 
Notably, it was found that some classes  of soft interactions lead 
to  reentrant melting transitions, others to polymorphic cluster phases~\cite{mladek2}, and in general to multiple  transitions 
involving close-packed and non-close-packed crystalline phases~\cite{suto,LikosNat,pephertz} as a function of the system density.  Remarkably,  the phase behavior of these systems is very much dependent on the 
the shape of the pair potential. 
Likos et al.~\cite{likos}  established a criterion 
to predict whether for a bounded and repulsive potential reentrant melting or cluster phases will occur based
on the sign of the Fourier transform of the interaction.
Nevertheless, there is currently no method to predict \textit{a priori} what specific crystal structures may become 
accessible for a given potential. 
For a recent review on the subject we refer the reader to reference~\cite{LikosRev}. 
    
In this paper we focus on two-dimensional systems.
Although much of the research in this field has focused on three-dimensional systems,
earlier numerical work on particles interacting via a hard core plus a shoulder potential
in two dimensions has revealed 
a phenomenological behavior that can be as rich as that observed at higher dimensionality
~\cite{Malescio,Stell,jagla,glaser}.  
Here we focus on  bounded soft potentials. Specifically, expanding on our recent results 
on Hertzian spheres~\cite{pephertz} and dumbbells~\cite{saric2} in three dimensions, we analyze the phase behavior of soft
elastic particles in two dimensions. We generalize our results to spherical surfaces,  and 
discuss the generalized Thomson problem for our soft potentials  by identifying novel polyhedral structures
representing minimal energy configurations formed by soft particles constrained on the surface of a sphere at different packing densities.
\edit{Although the term ``packing'' generally refers to hard-particle or weakly compressed systems, here it should be understood to be describing the ordering of highly compressed soft and deformable particles.}

\section*{Methods}
To study the phase behavior of soft nanoparticles in two dimensions we performed numerical simulations.
We considered systems of at least $N=1000$ spherical particles and used the 
standard Monte Carlo method \edit{with random initial configuration} in the NPT and NVT ensemble for the planar and spherical case respectively.
\edit{Results for two-dimensional systems were reproduced starting from a square-crystal intial configuration.}
All simulations where run for a minimum of $10^6$ iterations. Any two particles in our system interact
via a Hertz potential. The Hertz potential describes the elastic energy penalty associated with an axial compression
of two deformable spheres. Its functional form can be generalized as follows
 \begin{equation} \label{Hertz}
V(r)= 
\begin{cases}
  \varepsilon(1-\frac{r}{\sigma})^{\alpha} & \text{for  $r$ $\leq \sigma$}\cr
0 & \text{for  $r$ $ > \sigma$}\cr
\end{cases} 
\end{equation}
where $\sigma$ is the particle diameter, $r$ is the interparticle distance, $\varepsilon$ is the unit of energy, 
and $\alpha$ is a 
parameter that we control to modulate the shape of the interaction. The elastic
case is recovered for $\alpha=5/2$. 

\begin{figure}
	\includegraphics[width=0.5\textwidth]{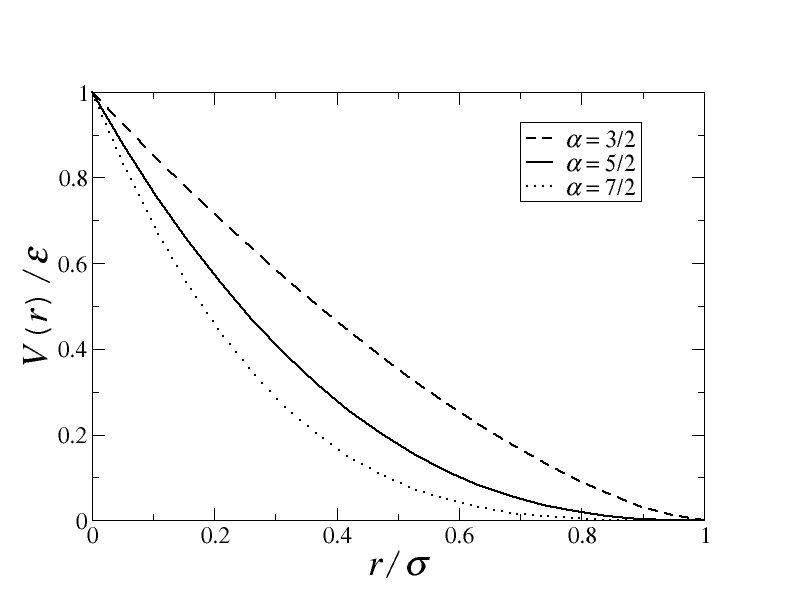}
	\caption{Plot of the rescaled  pair potential $V(r)/\varepsilon=(1-\frac{r}{\sigma})^{\alpha}$ 
	for the three values of $\alpha$ used in this study: $\alpha =\left\{\frac{3}{2}, \frac{5}{2}, \frac{7}{2}\right\}$}\label{pot}
\end{figure}

The Hertz model,  developed to account for small elastic deformations,
becomes inaccurate when associated to the large overlaps among particles that is
achieved at large densities; nevertheless, it provides us with a simple representation of a 
finite-ranged, bounded soft potential with a positive definite Fourier transform 
(a condition that guarantees reentrant melting in the phase diagram.) 
We included in our study two more values of $\alpha$ to account for a slightly harder 
($\alpha=3/2$) and a slightly weaker ($\alpha=7/2$) potential than the elastic one ($\alpha=5/2$.)
Figure~\ref{pot} shows these potentials. 
  
Crystallization in our simulations was determined using a combination of visual inspection and numerical order parameters similar to those described in~\cite{binder}. Specifically,  given a randomly chosen particle $j$, and a reference direction set to be along the bond between this particle and one of its nearest neighbors, we can define an order parameter sensitive to a crystal of $D$-fold symmetry (specifically six-fold hexagonal and four-fold square crystals) as follows:
For all particles $k$ within a radius of some cutoff $r_c$ of $j$ (including $j$ itself), we calculate the angle 
$\phi_{lk}$ between the bond joining each of the $D$ nearest neighbors $l$ of $k$ to $k$ and the reference direction.
If the number of particles within $r_c$ of $j$ is $N_k$, then the order parameter $\Psi_D^j$ is given by:
\edit{\[\Psi_D^j = \frac{1}{N_k}\frac{1}{D} \left|\sum_{k} \sum_{l} e^{i D \phi_{lk}}\right|\]}
This process is then repeated $N$ times and an average $\Psi_D$ is calculated, yielding a value between $0$ (for completely disordered systems) and $1$ (for perfectly crystalline systems).
This order parameter is analogous to that defined in~\cite{binder} except that instead of $\Psi_D^i$ being an average over all particles $k$ in the system, it is only over particles $k$ within $r_c$ ($= 1.65\sigma$ in our case); thus it is ``semi-local'' and allows crystalline order to be detected even in the presence of metastable grain boundaries.
The order parameter in~\cite{binder} represents the $r_c \rightarrow \infty$ limit of the above method.
 
To draw the phase diagrams we scan the phase space defined by the rescaled temperature 
$T_{\rm r}=k_{\rm B}T/\varepsilon$ and the rescaled number density $\rho_{\rm r}=N\sigma^2/A$ (where $A$ is the area of the system) to evaluate the several phases of the system in terms (when possible) of its order parameter 
$\Psi_D$.
\edit{The phase of the system was evaluated on a grid with a resolution of 0.001 in $T_{\rm r}$ and roughly 0.25 in $\rho_{\rm r}$.}
The lines separating the different regions are guides to the eye\edit{, and are placed between data points relative to structures having different symmetry}. 
Given the peculiarities of phase transitions in two dimensions, and the richness in the phases behavior reported, we have not attempted to establish equilibrium boundaries between the different phases, yet we
have reproduced our results using Molecular Dynamics simulations with a Langevin Thermostat performed in key regions of the phase space, and compared the stability of the crystalline structures by measuring their relative energies at $T_{\rm r}\rightarrow 0$.
\edit{Because thermodynamic phase diagrams are not presented, the phase diagrams that follow should be considered as semi-qualitative illustrations of the variety of phases which occur and the trend with which their occurrence follows the system density; specifics of the slopes and the precise location of the dividing lines between phases would require a more accurate study of the thermodynamic stability of the phases.}

\edit{For the simulations of the generalized Thomson problem in the latter half of the paper, a temperature quench was performed from a relatively high temperature ($T_r = 0.1$) to a very low temperature ($T_r = 10^{-7}$) in order to approach the zero-temperature limit.
For each $r_0$, the system was simulated with three initial configuration; one random, and one each from the structure obtained at slightly larger and slightly smaller $r_0$.}

\section*{Results}

\begin{figure}
 	\includegraphics[width=0.45\textwidth]{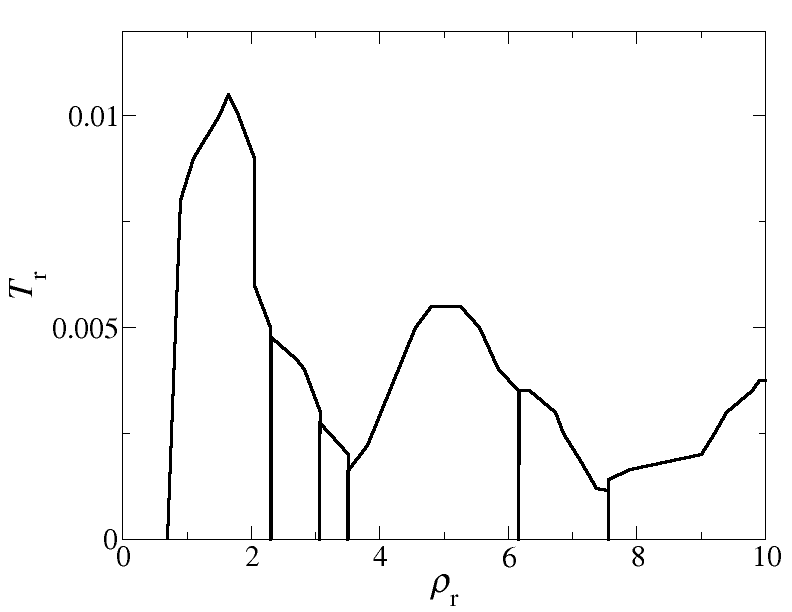}
		\label{fivehalvesdens}
		\put(-174,35){\subref*{triangular}}
		\put(-150,35){\subref*{square}}
		\put(-139,35){\subref*{pent}}
		\put(-110,35){\subref*{triangular}}
		\put(-74,35){\subref*{square}}
		\put(-33,35){\subref*{triangular}}
	\caption{Temperature vs. number density phase diagrams for two-dimensional particles interacting via a Hertz potential ($\alpha=5/2$). Labels on the plots refer to the labels of the phases in Fig.~\ref{phasep}.}\label{phaseHertz}  
\end{figure}
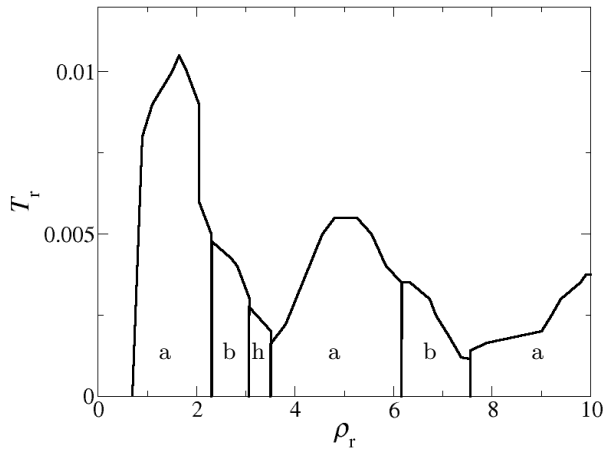

\begin{figure*}
	\subfloat[Hexagonal lattice]{
		\includegraphics[width=0.2\textwidth]{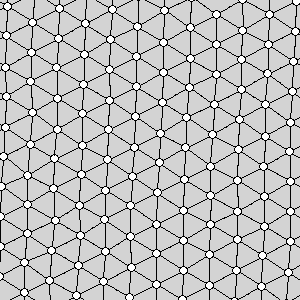}
		\label{triangular}
	}
	\subfloat[Square lattice]{
		\includegraphics[width=0.2\textwidth]{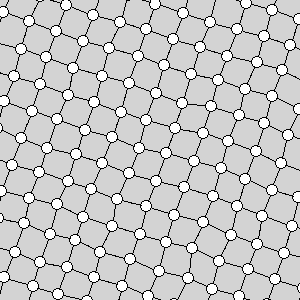}
		\label{square}

	} 
	\subfloat[Lines]{
		\includegraphics[width=0.2\textwidth]{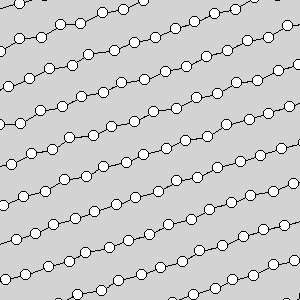}
		\label{line}
	}
	\subfloat[Stretched honeycomb lattice]{
		\includegraphics[width=0.2\textwidth]{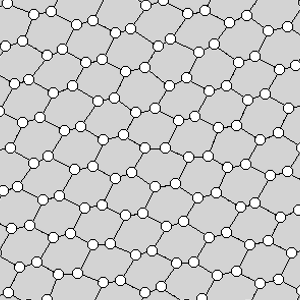}
		\label{honey}
	}\\ 
	\subfloat[Kagome lattice]{
		\includegraphics[width=0.2\textwidth]{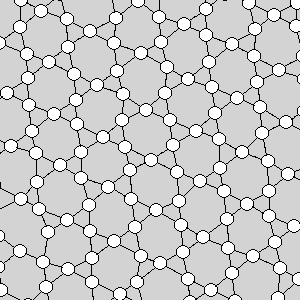}
		\label{kagome}
	}
	\subfloat[Open square lattice]{
		\includegraphics[width=0.2\textwidth]{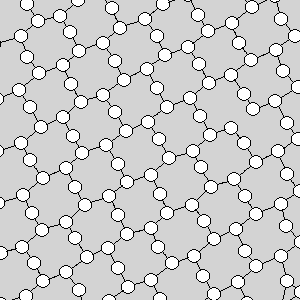}
		\label{largesquare}
	}
	\subfloat[Snub square tiling]{
		\includegraphics[width=0.2\textwidth]{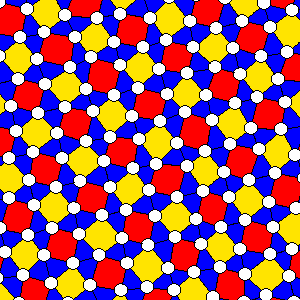}
		\label{snub}
	}
	\subfloat[Pentagons]{
		\includegraphics[width=0.2\textwidth]{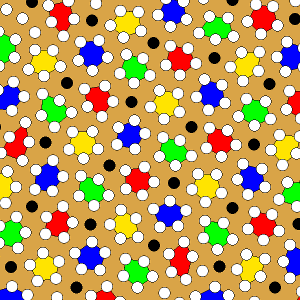}
		\label{pent}
	}
	\caption{Two-dimensional crystal phases formed in these systems.}\label{phasep}
\end{figure*}

We begin by presenting the phase diagram for the case of Hertzian (elastic) spheres.
The result is shown in Fig.~\ref{phaseHertz}.
The labels in the figure refer to the structures shown in Fig.~\ref{phasep}. 
As expected from Likos's criterion~\cite{likos} our system shows reentrant melting behavior
and a highest temperature above which the system remains fluid at all densities.
\edit{The main idea behind the formulation of the criterion is that at such large densities each particle is typically  interacting with such a 
significant  number of neighbors that the simple mean field approximation for the excess free energy of the system becomes quite accurate. This leads to a pair correlation function which is trivially proportional to the interaction potential $c(r)=-\beta v(r)$ and results in an analytical expression for the structure factor that has the form $S(q)=1/(1+\beta\rho\tilde{v}(q))$. This simple expression for $S(q)$ allows one to deduce several  properties of the fluid phase. For instance, to understand the presence of an upper freezing temperature or the occurrence of reentrant melting, one should realize that $S(q)$ cannot become singular (indicating the instability of the fluid phase ) if $\tilde{v}(q)$ (the Fourier transform of the potential) is positive defined. Since it is alway possible to find a sufficiently large temperature or density for which the mean field approximation becomes exact, one should expect the fluid phase to be stable under such conditions. 
Solid phases are  possible when the mean field approximation 
becomes inaccurate or for potential having a non-positive-defined Fourier transform. 
For more details on the subject we refer the reader to ref.~\cite{likos}.}

The phase behavior here reported is qualitatively similar to that observed in three dimensions, 
however the two-dimensional system does not produce a phase space that is as structurally rich  as its three dimensional counterpart. 
Across the spectrum of  densities (up to $\rho_{\rm r}=10$) and temperatures (down to 
$T_{\rm r}=10^{-6}$) that we explored, we find multiple crystal-to-crystal 
transitions from hexagonal (phase a) to square symmetry (phase b). 
Interestingly,  the occurrence of the two phases has an alternating periodic pattern, and no isostructural transition was detected in our system~\cite{mladek3}.
\edit{The multiple occurences of a single phase (for example, phases (a) and (b) in Fig.~\ref{phaseHertz}) are identical except for uniform scaling of the interparticle distances as the density of the system is changed.}

This pattern is only broken at a small range of densities centered around $\rho_{\rm r}\simeq 3.25$.
This region (phase h) is characterized by a complex structure dominated by particles arranged mostly into
pentagonal units. Our color coding of the phase shown in Fig.~\ref{phasep}(h) helps rationalizing the overall
symmetry of the structure which in this representation can be thought of as the combination of four 
square lattices. Three lattices are generated by pentagonal units, displaced with
respect to each other but  oriented along the same axis. The fourth lattice is built out
 of the non-connected particles and is oriented along an axis 
 that forms an angle of 45 degrees with the other lattices. 

\begin{figure}
	\subfloat{
		\includegraphics[width=0.45\textwidth]{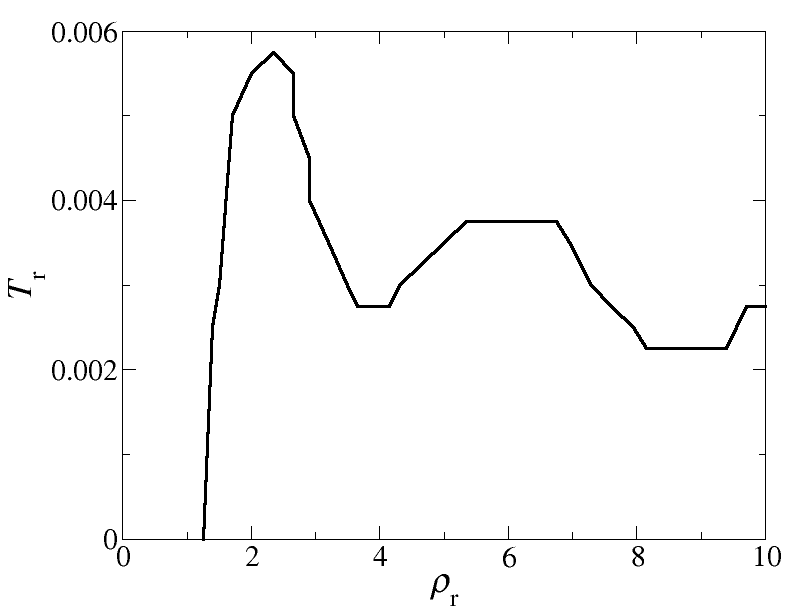}
		\label{sevenhalvesdens}
		\put(-104,60){\subref*{triangular}}
	}\\
	\subfloat{	
	\includegraphics[width=0.45\textwidth]{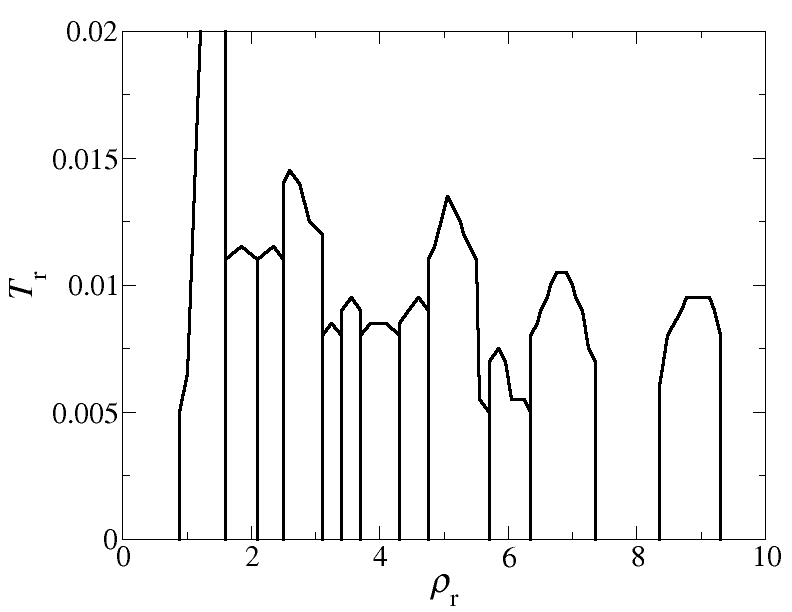}
	\label{threehalvesdens}
	\put(-177,40){\subref*{triangular}}
	\put(-166,40){\subref*{square}}
	\put(-157,40){\subref*{line}}
	\put(-148,40){\subref*{honey}}
	\put(-138.5,40){\subref*{kagome}}
	\put(-133,40){\subref*{largesquare}} 
	\put(-125,40){\subref*{snub}}
	\put(-115,40){\subref*{triangular}}
	\put(-103,40){\subref*{square}}
	\put(-88,40){\subref*{pent}}
	\put(-73,40){\subref*{honey}}
	\put(-34,40){\subref*{triangular}}	
	}
	\caption{Temperature vs. number density phase diagrams for two-dimensional particles interacting via   the generalized Hertz potential  with $\alpha =\frac{7}{2}$ (top) and  $\alpha=\frac{3}{2}$(bottom).  
Labels on the plots refer to the labels of the phases in Fig.~\ref{phasep}.}\label{phased} 	
\end{figure}
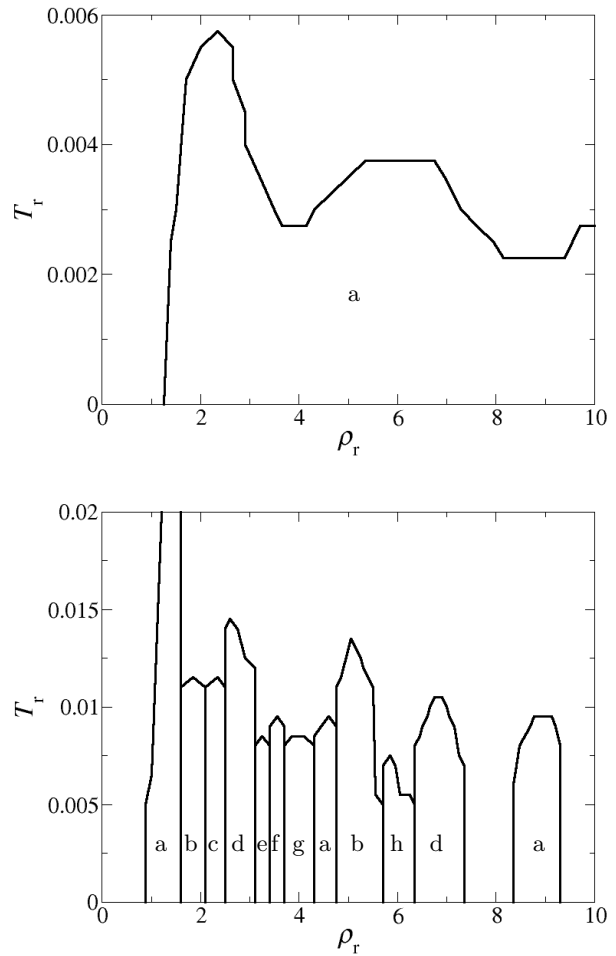

To understand how the phase behavior is affected by the specific choice of the functional 
form of the potential, we repeated the calculation of the phase diagram for 
$\alpha=7/2$ and $\alpha=3/2$.
The results are presented is Fig.~\ref{phased}. Although the overall
trend is very similar, the number and the sequence of phases that we find are significantly different.
Namely, $\alpha=3/2$ leads to a significantly richer structural behavior, while the 
large value of $\alpha$ results in a single hexagonal crystalline structure.
As a reference, notice  that our potential tends  to the linear-ramp potential for $\alpha=1$
(for which a large number of phases including quasi-crystals have been observed~\cite{jagla})
, we recover the square-shoulder potential for $\alpha\rightarrow 0$
(for which a cascade of cluster phases are expected), and tends to a Kronecker delta potential
for  $\alpha\rightarrow\infty$ (for which no crystallization is expected at any finite density).

\begin{figure}
\includegraphics[width=0.45\textwidth]{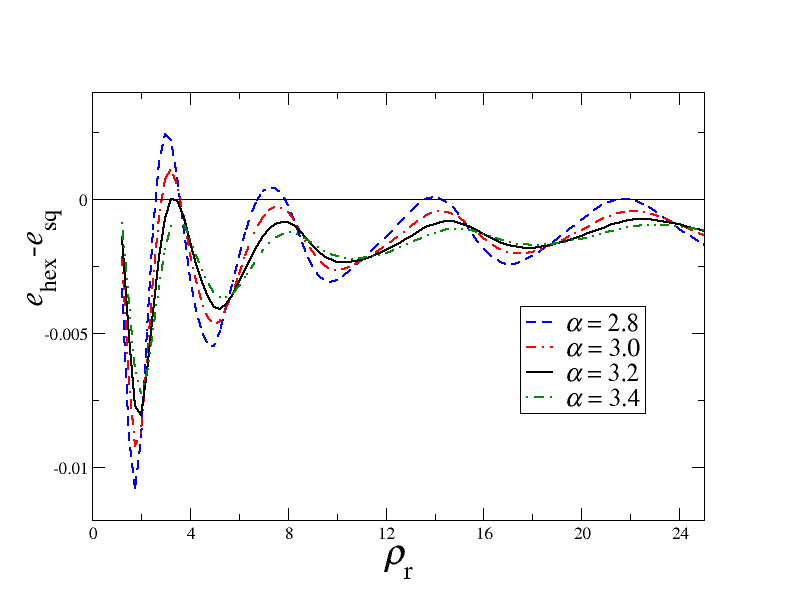}
\caption{Energy density difference between hexagonal and square lattice at $T_{\rm r}=0$ as a 
function of system density at different values of $\alpha$. We estimated the 
critical $\alpha$ to be around $\alpha\simeq 3.1822$.}
\label{energies}
\end{figure}

At least as a general trend, it is therefore not surprising that structural 
variety increases for smaller values of $\alpha$. In this regard, it is interesting to estimate
the critical power above which the hexagonal lattice becomes the only stable lattice as observed 
in Fig.~\ref{phased}(bottom).
Given that the only other competing structure at large values of $\alpha$ is the square lattice, we
computed the energy density difference between hexagonal and square lattices as a function of
density and at zero temperature for different values of $\alpha$. The result is presented
 in Fig.~\ref{energies} for densities up to $\rho_{\rm r}=25$ (our full numerical data extend up
 to $\rho_{\rm r}=100$, but are not shown for the sake of clarity).
We find that the hexagonal lattice becomes more stable than the square lattice at all 
densities for any powers above the onset value $\alpha^*\simeq 3.182$. 
Several numerical simulations at finite densities for powers ranging from $\alpha=10$ to $\alpha=50$ 
where carried out, and indeed only hexagonal structures where observed upon system ordering.
The presence of reentrant melting was also established for $\alpha=5$ and $\alpha=10$.
The oscillating behavior  of the energy density difference for $\alpha<\alpha^*$ also explains 
qualitatively the periodic structural pattern observed in the phase diagram for $\alpha=5/2$. 
\edit{The precise underlying physical origin of this periodic structural pattern is currently under further investigation.
Early results indicate that the finite range of the potential is a significant contributor to the behavior; in particular, the fact that as the density of the system increases, the number of particles with which each particle in the system can interact increases in a stepwise fashion.}

We now turn to the discussion of the second part of the paper. Given the structural variety
observed in the two dimensional compression of soft nanoparticles, it is interesting to 
extend our results to other geometries. The spherical case is of particular interest 
because a large body of work has been dedicated to finding and enumerating minimal 
energy conformations of particles constrained over a spherical shell, and 
repelling each other via an electrostatic repulsion ~\cite{altschuler, erber, perez-garrido, morris, erber2, edmundson, altschuler2, dodgson, perez-garrido2, perez-garrido3,Bowick2,Bowick3}.
This is what Thomson asked in 1904~\cite{thomson}  when attempting to construct his plum pudding
model of the atom. Although there is consensus on the minimal energy structures for
$N<100$, the problem becomes more involved for large values of $N$ for which an exponentially large
number of low energy configurations becomes available.  In this regime, lattices with overall icosahedral
symmetry ({\it icosadeltahedra}) have been initially postulated as possible global minima for
specific magic number of particles~\cite{altschuler2},  but eventually 
it was shown that it is possible to lower the energy of these configurations adding dislocation 
defects~\cite{perez-garrido3,Bowick}. Grain boundary scars have also been observed in
experiments~\cite{bausch}. 
 
\begin{figure*}
	\subfloat{
	\includegraphics[width=0.25\textwidth]{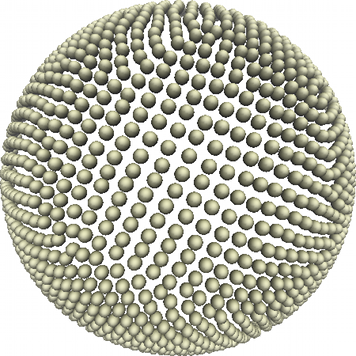}
	} 
	\subfloat{	
	\includegraphics[width=0.25\textwidth]{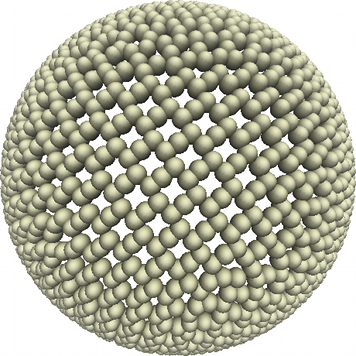}
	}
	\subfloat{	
	\includegraphics[width=0.25\textwidth]{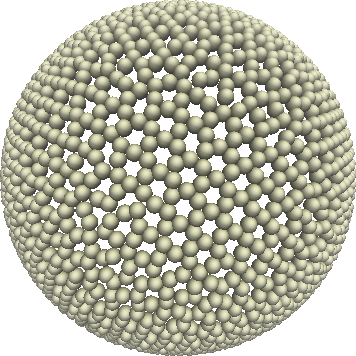}
	}
\caption{Snapshots of soft particles packing on a spherical surface.
Here we show from left to right square, open square and kagome lattices.}\label{spheres}	
\end{figure*}

Although we confirm that the same pattern of phases obtained in two dimensions 
also develops on large areas of spherical shells -- apart from the topologically required defects --
as a function of  the radius of the sphere (see Fig.~\ref{spheres} for a sample of these configurations), we have not attempted to 
enumerate all possible low energy states for large $N$. It is however clear that considering
soft potentials introduces two new parameters to the system: namely the packing density and the 
shape of the potential. A careful analysis of the defects and symmetries arising in this 
regime is currently under investigation and will be the subject of  another publication; here we focus on
the case of small number of particles $N\leq12$ for $\alpha=3/2$ and $\alpha=5/2$.

\begin{figure*}
\scriptsize\begin{tabular}{>{\centering}m{12px} | >{\centering}m{0.14\textwidth} >{\centering}m{0.14\textwidth} >{\centering}m{0.14\textwidth} >{\centering}m{0.14\textwidth} >{\centering}m{0.14\textwidth} >{\centering}m{0.14\textwidth} >{\centering}m{0px} }
$N_b$ & \multicolumn{6}{c}{Cluster geometry} \\
\hline
\multirow{4}{*}{5} & \includegraphics[width=0.10\textwidth]{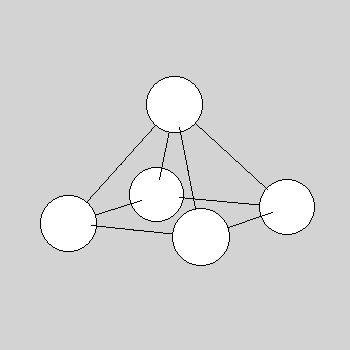} & \includegraphics[width=0.10\textwidth]{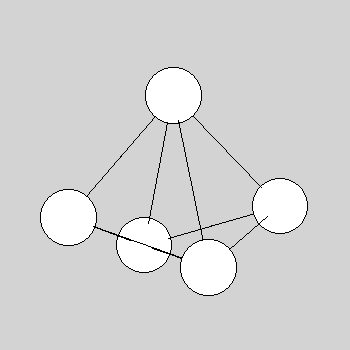} &  \inbox{\includegraphics[width=0.10\textwidth]{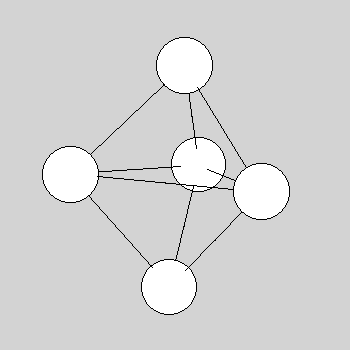}} & & &\\
& square pyramid & $P_{\rm A}^{(5)}$ & \inbox{trigonal dipyramid} & & & &\\
& $r_0 = 0.65$ & $r_0 = 0.58$ & \inbox{$r_0 = 0.50$} & & & &\\
& $E = 0.116675$ & $E = 0.457195$ & \inbox{$E = 1.098194$} & & & &\\\cline{4-4}\\[-7pt]

\cline{3-3}
& & \inbox{} & & & & &\\[-7pt]
\multirow{4}{*}{7} & \includegraphics[width=0.10\textwidth]{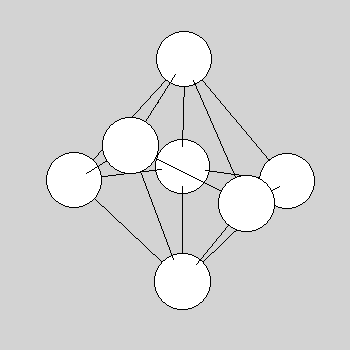} & \inbox{\includegraphics[width=0.10\textwidth]{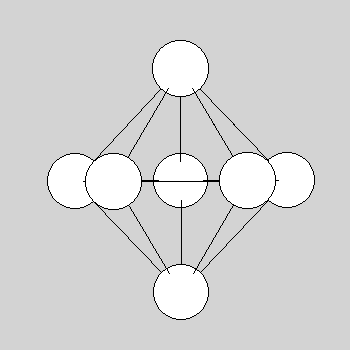}} & \includegraphics[width=0.10\textwidth]{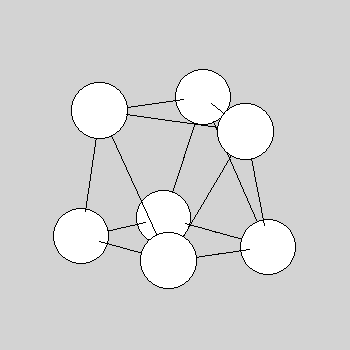} & \includegraphics[width=0.10\textwidth]{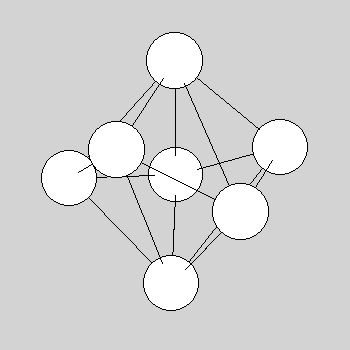} & \includegraphics[width=0.10\textwidth]{7c.png} & \includegraphics[width=0.10\textwidth]{7a.png}& \\
&  $P_{\rm A}^{(7)}$& \inbox{pentagonal dipyramid} &$P_{\rm B}^{(7)}$ &$P_{\rm C}^{(7)}$ &$P_{\rm B}^{(7)}$ &$P_{\rm A}^{(7)}$&\\
& $r_0 = 0.75$ & \inbox{$r_0 = 0.65$} & $r_0 = 0.58$ & $r_0 = 0.55$ & $r_0 = 0.50$ & $r_0 = 0.40$ & \\
& $E = 0.1481239$ & \inbox{$E = 0.802319$} & $E = 1.635489$ & $E = 2.066014$ & $E = 2.956885$ & $E = 5.463620$ & \\ \cline{3-3}\\[-7pt]

\cline{2-2}\cline{4-4}
& \inbox{} & & \inbox{} & & & &\\[-7pt]
\multirow{4}{*}{8} &\inbox{\includegraphics[width=0.10\textwidth]{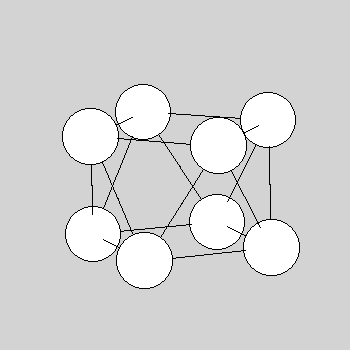}} & \includegraphics[width=0.10\textwidth]{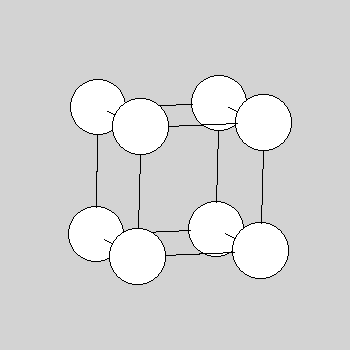} & \inbox{\includegraphics[width=0.10\textwidth]{8a.png}} & & & & \\
& \inbox{square antiprism} & cube & \inbox{square antiprism} & & &\\
& \inbox{$r_0 = 0.70$} & $r_0 = 0.58$ & \inbox{$r_0 = 0.50$} & & & & \\
& \inbox{$E = 0.851626$} & $E = 2.423538$ & \inbox{$E = 4.212084$} & & & & \\\cline{2-2}\cline{4-4}\\[-7pt]

\cline{2-2}\cline{4-4}\cline{7-7}
& \inbox{} & & \inbox{} & & & \inbox{} &\\[-7pt]
\multirow{4}{*}{9} & \inbox{\includegraphics[width=0.10\textwidth]{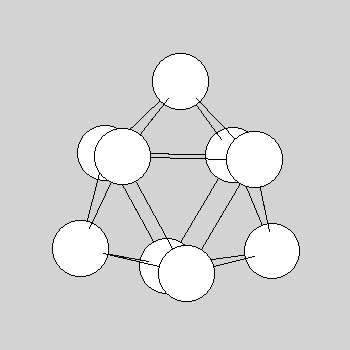}} & \includegraphics[width=0.10\textwidth]{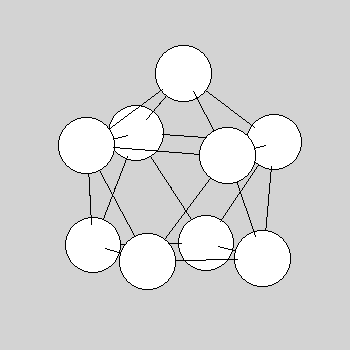} &  \inbox{\includegraphics[width=0.10\textwidth]{9a.png}} & \includegraphics[width=0.10\textwidth]{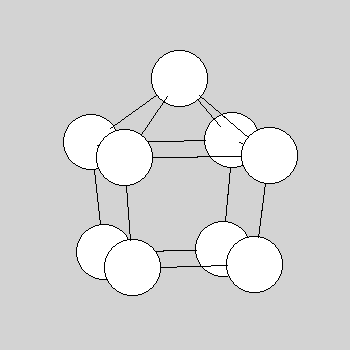} & \includegraphics[width=0.10\textwidth]{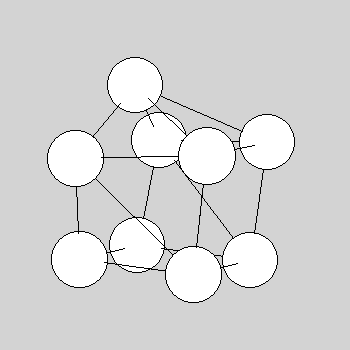} & \inbox{\includegraphics[width=0.10\textwidth]{9a.png}} & \\
& \inbox{triaugmented triangular prism} & gyroelongated square pyramid  & \inbox{triaugmented triangular prism} & elongated square pyramid & $P_{\rm A}^{(9)}$ & \inbox{triaugmented triangular prism} &\\
& \inbox{$r_0 = 0.81$} & $r_0 = 0.78$ & \inbox{$r_0 = 0.70$} & $r_0 = 0.63$ & $r_0 = 0.58$ & \inbox{$r_0 = 0.50$} & \\
& \inbox{$E = 0.268064$} & $E = 0.508830$ & \inbox{$E = 1.424194$} & $E = 2.489422$ & $E = 3.479048$ & \inbox{$E = 5.696124$} & \\ \cline{2-2} \cline{4-4} \cline{7-7} \\[-7pt]

\cline{5-5}
& & & & \inbox{} & & &\\[-7pt]
\multirow{4}{*}{10} & \includegraphics[width=0.10\textwidth]{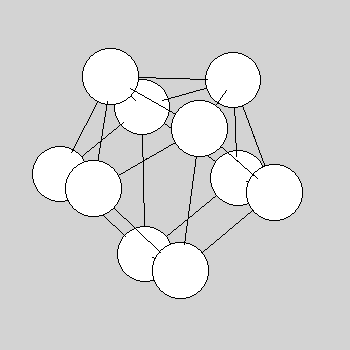} & \includegraphics[width=0.10\textwidth]{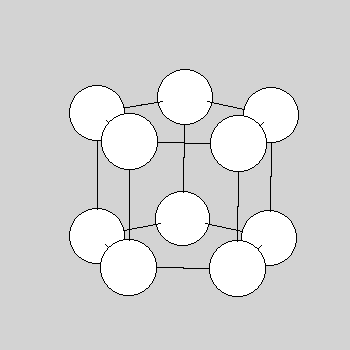} & \includegraphics[width=0.10\textwidth]{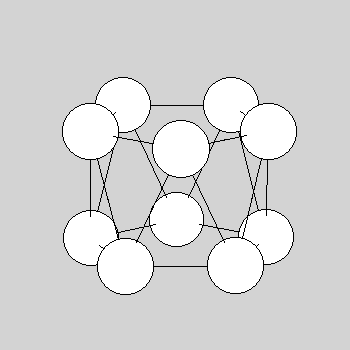} & \inbox{\includegraphics[width=0.10\textwidth]{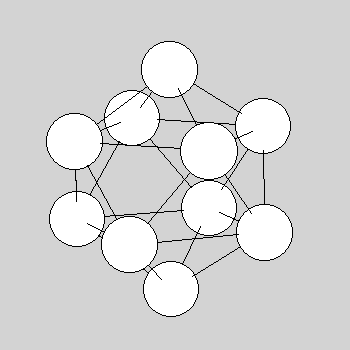}} & & & \\
& sphenocorona & pentagonal prism & pentagonal antiprism & \inbox{gyroelongated square dipyramid} & &\\
& $r_0 = 0.90$ & $r_0 = 0.60$ & $r_0 = 0.57$ & \inbox{$r_0 = 0.40$} & & & \\
& $E = 0.035971$ & $E = 4.102489$ & $E = 4.921405$ & \inbox{$E = 12.689921$} & & & \\ \cline{5-5}\\[-7pt]

\cline{7-7}
& & & & & & \inbox{} &\\[-7pt]
\multirow{4}{*}{11} & \includegraphics[width=0.10\textwidth]{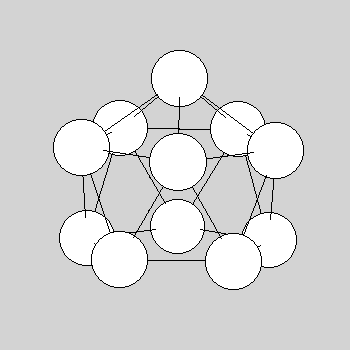} & \includegraphics[width=0.10\textwidth]{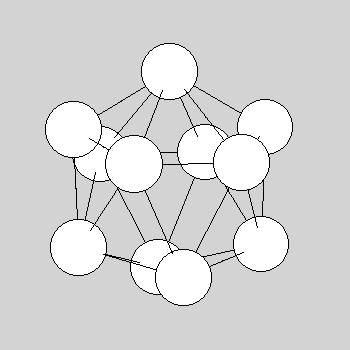} & \includegraphics[width=0.10\textwidth]{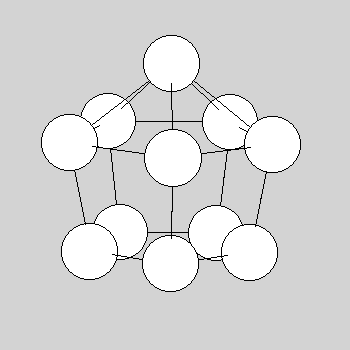} & \includegraphics[width=0.10\textwidth]{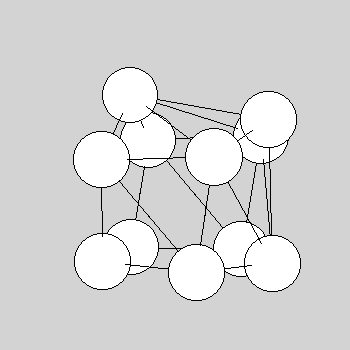} & \includegraphics[width=0.10\textwidth]{11a.png} & \inbox{\includegraphics[width=0.10\textwidth]{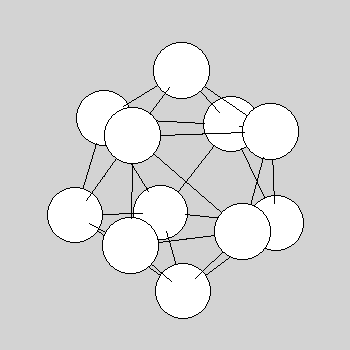}} & \\
& gyroelongated pentagonal pyramid &$P_{\rm A}^{(11)}$  & elongated pentagonal pyramid &$P_{\rm B}^{(11)}$ & gyrelongated pentagonal pyramid & \inbox{$P_{\rm C}^{(11)}$} &\\
& $r_0 = 0.90$ & $r_0 = 0.80$ & $r_0 = 0.68$ & $r_0 = 0.63$ & $r_0 = 0.55$ & \inbox{$r_0 = 0.40$} & \\
& $E = 0.244728$ & $E = 1.308288$ & $E = 3.322424$ & $E = 4.479854$ & $E = 7.042131$ & \inbox{$E = 15.783736$} & \\ \cline{7-7}\\[-7pt]

\cline{2-2} \cline{6-6}
& \inbox{} & & & & \inbox{} & &\\[-7pt]
\multirow{4}{*}{12} & \inbox{\includegraphics[width=0.10\textwidth]{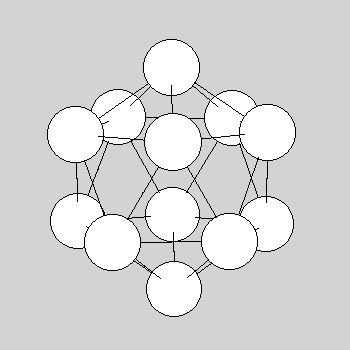}} & \includegraphics[width=0.10\textwidth]{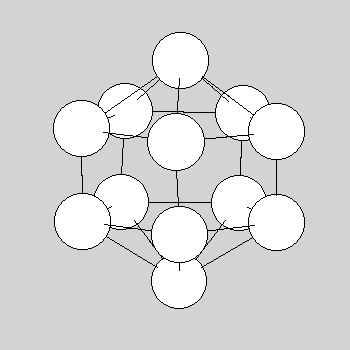} & \includegraphics[width=0.10\textwidth]{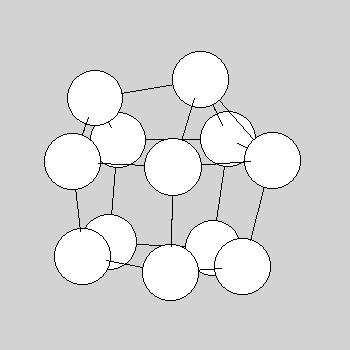} & \includegraphics[width=0.10\textwidth]{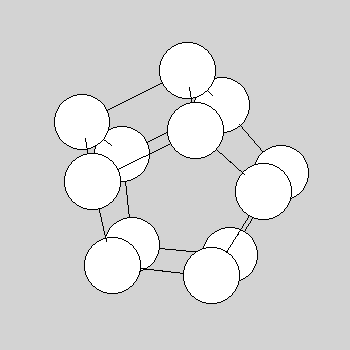} & \inbox{\includegraphics[width=0.10\textwidth]{12a.png}} & & \\
& \inbox{icosahedron} & elongated pentagonal dipyramid &$P_{\rm A}^{(12)}$ &$P_{\rm B}^{(12)}$ & \inbox{icosahedron} &\\
& \inbox{$r_0 = 0.90$} & $r_0 = 0.75$ & $r_0 = 0.70$ & $r_0 = 0.65$ & \inbox{$r_0 = 0.50$} & & \\
& \inbox{$E = 0.373165$} & $E = 2.857171$ & $E = 3.806734$ & $E = 5.087802$ & \inbox{$E = 11.529991$} & & \\ \cline{2-2} \cline{6-6}

\end{tabular}
\caption{Global minimum configurations formed by small number of particles constrained on the surface of a sphere interacting
with a generalized hertzian soft potential for $\alpha=3/2$, at different values of reduced sphere radius $r_0$. The reduced energy of the configurations is also indicated together with the name of the polyhedric structures. Some structures for which no name
was found are indicated with the nomenclature  $P^{(N)}_{\rm X}$. For $N=4$ and $N=6$ (not shown) we obtain respectively 
a tetrahedral and an octahedral structure for every $r_0$. Reference 
configurations for long-range potentials have been highlighted with a dark border.}\label{clusters1}
\end{figure*}

\begin{figure*}
\scriptsize\begin{tabular}{>{\centering}m{12px} | >{\centering}m{0.14\textwidth} >{\centering}m{0.14\textwidth} >{\centering}m{0.14\textwidth} >{\centering}m{0.14\textwidth} >{\centering}m{0.14\textwidth} >{\centering}m{0.14\textwidth} >{\centering}m{0px} }
$N_b$ & \multicolumn{6}{c}{Cluster geometry} \\
\hline
\multirow{4}{*}{5} & \includegraphics[width=0.10\textwidth]{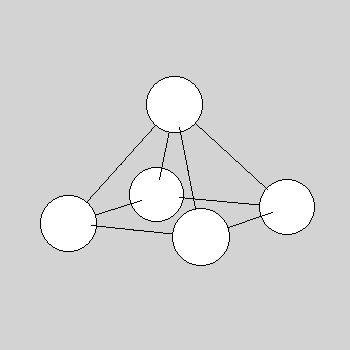} & \includegraphics[width=0.10\textwidth]{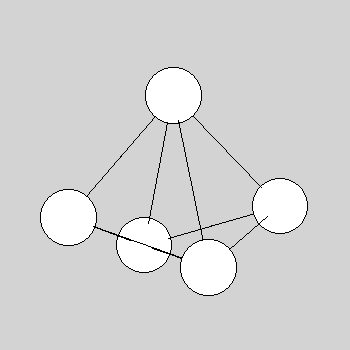} &  \inbox{\includegraphics[width=0.10\textwidth]{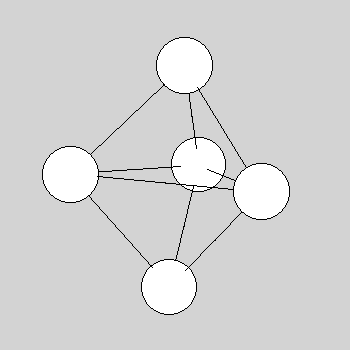}} & & &\\
& square pyramid & $P_{\rm A}^{(5)}$& \inbox{trigonal dipyramid} & & & &\\
& $r_0 = 0.70$ & $r_0 = 0.58$ & \inbox{$r_0 = 0.50$} & & & &\\
& $E = 0.000047$ & $E = 0.081254$ & \inbox{$E = 0.298277$} & & & &\\ \cline{4-4}\\[-7pt]

\cline{3-3}
& & \inbox{} & & & & &\\[-7pt]
\multirow{4}{*}{7} & \includegraphics[width=0.10\textwidth]{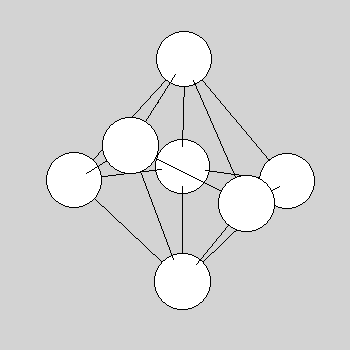} & \inbox{\includegraphics[width=0.10\textwidth]{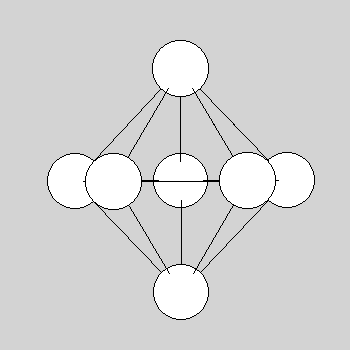}} & \includegraphics[width=0.10\textwidth]{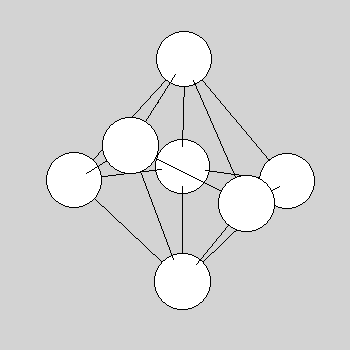} & & & & \\
&$P_{\rm A}^{(7)}$ & \inbox{pentagonal dipyramid} & $P_{\rm A}^{(7)}$& & &\\
& $r_0 = 0.75$ & \inbox{$r_0 = 0.50$} & $r_0 = 0.25$ & & & & \\
& $E = 0.008832$ & \inbox{$E = 1.012408$} & $E = 6.627468$ & & & & \\\cline{3-3}\\[-7pt]

\cline{2-2} \cline{4-4}
& \inbox{} & & \inbox{} & & & &\\[-7pt]
\multirow{4}{*}{9} & \inbox{\includegraphics[width=0.10\textwidth]{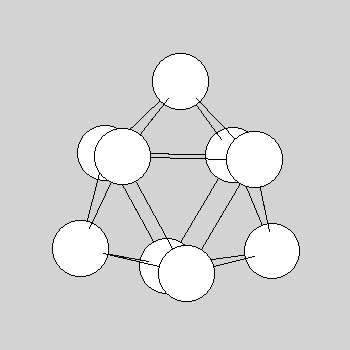}} & \includegraphics[width=0.10\textwidth]{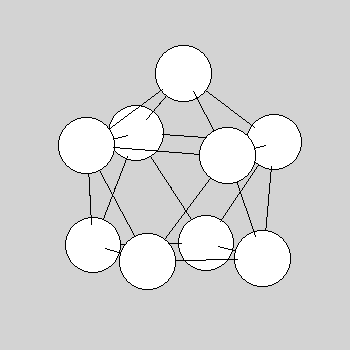} & \inbox{\includegraphics[width=0.10\textwidth]{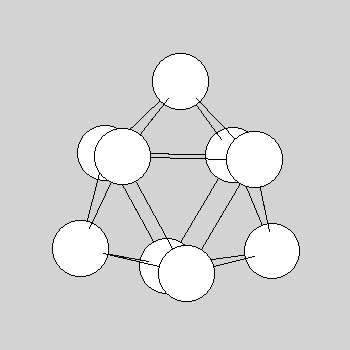}} & & & & \\
& \inbox{triaugmented triangular prism} & gyroelongated square pyramid & \inbox{triaugmented triangular prism} & & & &\\
& \inbox{$r_0 = 0.75$} & $r_0 = 0.72$ & \inbox{$r_0 = 0.50$} & & & & \\
& \inbox{$E = 0.110712$} & $E = 0.196371$ & \inbox{$E = 2.203406$} & & & & \\\cline{2-2}\cline{4-4}\\[-7pt]

\cline{3-3} \cline{5-5}
& & \inbox{} & & \inbox{} & & &\\[-7pt]
\multirow{4}{*}{10} & \includegraphics[width=0.10\textwidth]{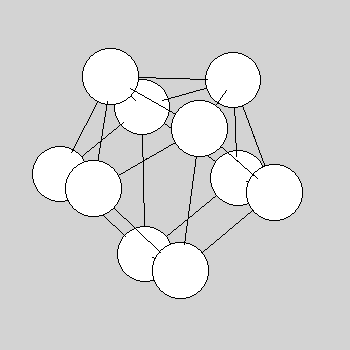} & \inbox{\includegraphics[width=0.10\textwidth]{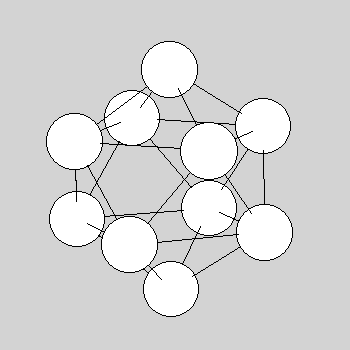}} & \includegraphics[width=0.10\textwidth]{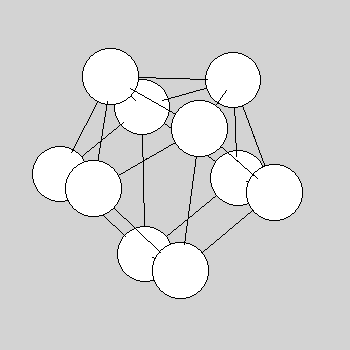} & \inbox{\includegraphics[width=0.10\textwidth]{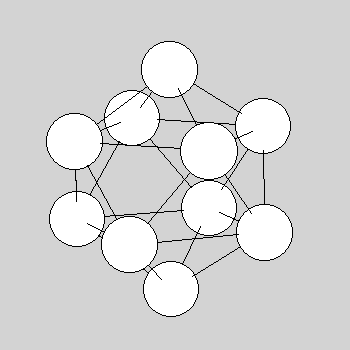}} & & & \\
& sphenocorona & \inbox{gyroelongated square dipyramid} & sphenocorona & \inbox{gyroelongated square dipyramid} & &\\
& $r_0 = 0.80$ & \inbox{$r_0 = 0.65$} & $r_0 = 0.60$ & \inbox{$r_0 = 0.50$} & & & \\
& $E = 0.097219$ & \inbox{$E = 0.850254$} & $E = 1.361079$ & \inbox{$E = 2.986067$} & & & \\\cline{3-3}\cline{5-5}\\[-7pt]

\cline{3-3}
& & \inbox{} & & & & &\\[-7pt]
\multirow{4}{*}{11} & \includegraphics[width=0.10\textwidth]{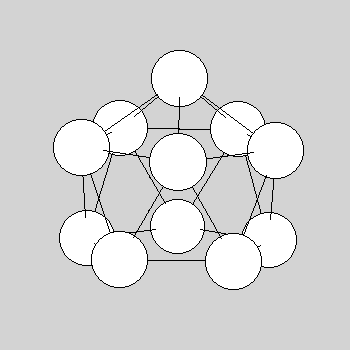} & \inbox{\includegraphics[width=0.10\textwidth]{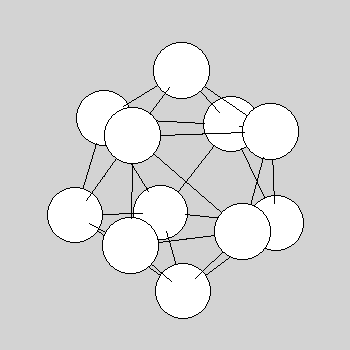}} & & & & & \\
& gyroelongated pentagonal pyramid & \inbox{$P_{\rm A}^{(11)}$} & & & &\\
& $r_0 = 0.80$ & \inbox{$r_0 = 0.70$} & & & & & \\
& $E = 0.221753$ & \inbox{$E = 0.793120$} & & & & & \\\cline{3-3}

\end{tabular}
\caption{Global minima configurations formed by small number of particles constrained on the surface of a sphere interacting
with a generalized hertzian soft potential for $\alpha=5/2$, at different values of reduced sphere radius $r_0$. The reduced energy of the configurations is also indicated together with the name of the polyhedric structures. Some structures for which no name
was found are indicated with the nomenclature  $P^{(N)}_{\rm X}$.
For $N=4$, $N=6$ and $N=12$ (not shown) we obtain respectively 
a tetrahedral, an octahedral and an icosahedral structure for every $r_0$. Reference 
configurations for long-range potentials have been highlighted with a dark border.
}\label{clusters2}
\end{figure*} 

Figures~\ref{clusters1} and \ref{clusters2}  show the results of our analysis for $N\in\{5,12\}$.
The rescaled configurational energy $E$ defined as the total internal energy divided by  
$\varepsilon$, the reduced radius $r_0=R/\sigma$ of the spherical surface and the 
names of the polyhedra (when available) are also given. 
The reference configurations for the electrostatic problem are highlighted with a dark frame.
These data are obtained using both Monte Carlo and Molecular Dynamics simulations 
with a standard temperature annealing procedure down to $T_{\rm r}\rightarrow 0$.
\edit{For each $r_0$ shown in the table, the depicted configuration was reached from at least three separate starting configurations, indicating that competing metastable states are not a major concern.
Note that Figures~\ref{clusters1} and \ref{clusters2} do no represent a complete enumeration of every cluster symmetry that may be obtained.  
They show the subset of configuration which are stable over some range of $r_0$; regions in which the particle configuration changes continuously with $r_0$ have not been tabulated.}

Unlike the case of charged point particles, we find that indeed 
the number and the nature of the structures is very much dependent on the radius of 
the sphere. The overall trend follows what found in the analysis of the planar two-dimensional system;
namely, smaller values of $\alpha$ lead to richer structural diversity.
In all cases we where able to recover the global minima of the electrostatic system for
at least one spherical radius. Interestingly, $N=4$ and $N=6$ are the only cases 
in which for both powers we obtain only one structure, 
a tetrahedron and a octahedron respectively (not shown), 
which are stable for any value of $r_0$ explored (down to $r_0=0.10$). 
The stability of the octahedron  ($N=6$) was also confirmed 
for values of $\alpha$ up to 10. Low energy configurations for values of $N$ larger than 12 
have also been considered, however, given the large variety and the complexity of the structures
arising upon increasing $N$, we have not attempted to enumerate them. In fact, 
more sophisticated algorithms than the one used in this work would be necessary to thoroughly
explore the energy landscape in search of a global minimum.

\section*{Conclusions}
In this paper we explore the phase behavior of \edit{model} colloidal particles interacting via a generalized soft (Hertz) potential.
Specifically, we analyze how structural diversity depends on the particular choice of the functional form of the potential
for particles constrained on a two-dimensional plane and on the surface of a spherical shell.
For the planar case we compute how the  phase diagram of the system changes with the functional form of 
the pair potential, and establish a limit above which (for the class of potentials explored) hexagonal packing becomes the only 
allowed symmetry of the ordered phase.
For small number of particles ($N\leq 12$), we identify on a spherical shell the polyhedral 
configurations establishing global energy minima
for different pair potentials, and compare the results to the global minima 
corresponding to the classic Thomson problem. We show that unlike the electrostatic case, for the same number of 
particles the system presents more than one global minimum depending on the radius of the sphere.

It would be interesting to study the geometry of the defects arising for larger number of particles on the spherical shell,
analyze the overall symmetry of the ground states, and explore 
whether patterns of grain boundaries, expected for large values of $N$ in the electrostatic system, would also
manifest for soft potentials. This is a very challenging problem that we have already begun to investigate and requires more sophisticated 
minimization techniques than the ones employed in this study.
For the time being, we have established that the same phases observed in the planar geometry do develop on the spherical shell 
at comparable number densities.
   
 Finally, given the sensitivity of the phase behavior on the specific choice of the pair potential, 
 it becomes critical at this point to develop sophisticated and reliable coarse-graining procedures able 
 to capture accurate interactions between dendrimers or star polymers, but also  the inverse problem, i.e. 
 the design of polymeric structures able to reproduce desired functional forms for the pair potential, is in great need 
 of a better theoretical understanding.
\section*{ACKNOWLEDGMENTS}
This work was supported by the American Chemical Society under PRF grant No. 50221-DNI6
A. C. thanks Vincent Nguyen for helping with the simulations.

\end{document}